\documentclass{nature}

\usepackage{graphicx}
\usepackage{amsmath}
\usepackage{amssymb}
\usepackage{color}
\usepackage{dcolumn}
\usepackage{epsfig}
\usepackage{bm}

\title{Crossover of the pairing symmetry from $s$- to $d$-wave  in iron pnictide superconductors}

\author{Mahmoud Abdel-Hafiez$^{1}$, Zheng He$^{1,2}$, Jun Zhao$^{2,3}$, Xingye Lu$^4$, Huiqian Luo$^4$, Pengcheng Dai$^{4,5}$ \& Xiao-Jia Chen$^1$}

\begin{document}

\maketitle

\begin{affiliations}
 \item Center for High Pressure Science and Technology Advanced Research, Shanghai, 201203, China
 \item State Key Laboratory of Surface Physics and Department of Physics, Fudan University, Shanghai 200433, China
 \item Collaborative Innovation Center of Advanced Microstructures, Nanjing 210093, China
 \item Beijing National Laboratory for Condensed Matter Physics, Institute of Physics, Chinese Academy of Sciences, Beijing 100190, China
 \item Department of Physics and Astronomy, Rice University, Houston, Texas 77005, USA

\end{affiliations}

\begin{abstract}
One of the major themes in the physics of condensed matter is unconventional superconductivity in correlated electronic systems. The symmetry structure of Cooper pairs is thought to be the key to the understanding of the pairing mechanism of their superconductivity\cite{Cup,SC}. In iron pnictide superconductors, the precise symmetry of the order parameter as well its evolution with doping remains highly controversial\cite{Hir}. There is no general consensus on the nature of pairing in iron-based superconductors leaving the perspectives ranging from $s^{++}$ wave, to $s^{\pm}$, and $d$ wave. Here we show that doping induced a sudden change of the pairing symmetry from the $s^{++}$- to $d$-wave in the BaFe$_{2-x}$Ni$_{x}$As$_{2}$ system with an $s^{\pm}$ state near the crossover boundary at $x=0.15$. The presence of the $s^{\pm}$ state is of particular scientific interest because an intermediate phase breaks the pairing symmetry\cite{s0,s1,s2,s3,s4,s5} when the $s$- and $d$-wave phases come together to interfere. For $x <$ 0.15, the temperature dependence of the London penetration depth, lower critical field, and specific heat in the superconducting (SC) state provides strong evidence for a two-band $s$-wave order parameter. Upon doping for $x >$ 0.15, we find the temperature and magnetic-field contributions to the specific heat in $T^{2}$ and $\sqrt{H}$, respectively, indicating line nodes in the SC gap. These findings thus clarify the symmetry nature as well as the evolution of the SC order parameter with doping in pnictide superconductors.
\end{abstract}

Conventional phonon-mediated superconductors and unconventional cuprate superconductors\cite{Cup} are well characterized by distinct $s$-wave and $d$-wave pairing symmetries with nodeless and nodal  gap distributions, respectively. In conventional superconductors, the electron-phonon interaction gives rise to the attraction between electrons near the Fermi-surface (FS) with opposite momenta and opposite spins, which eventually causes superconductivity. On the other hand, it has generally been believed that the electron-electron coupling leads to the formation of Cooper pairs in cuprates and pnictides\cite{Hir}. However, iron pnictides present a rich phase diagram wherein superconductivity coexists and competes with the spin density wave, resulting in unconventional pairing mechanisms\cite{d}. Although there is a general consensus that spin fluctuations play an important role in the formation of Cooper pairs, many aspects such as the role of magnetism, the nature of chemical tuning, and the resultant pairing symmetry remain unsettled\cite{SC,Hir}. In a $d$-wave pairing symmetry, the order parameter changes sign with angles in the $x$-$y$ plane, forcing the gap to go to zero along diagonal directions ($k_{y}=\pm k_{x}$)\cite{J}. In the case of $s$-wave symmetry, the gap in general will not have nodes. In the momentum space, an $s$-wave SC gap has an isotropic magnitude in all directions, and it has a fixed phase in all directions (see the inset of Fig.\,1). More specifically, it is additionally experimentally shown that the electronic specific heat, angle-resolved photoemission spectroscopy (ARPES), and scanning tunneling micrposcopy (STM) studies of Ba(Fe$_{1-x}$Co$_{x}$)$_{2}$As$_{2}$ agree on an isotropic $s$-wave pairing symmetry\cite{Hardy,ARPES,STM}, while Raman scattering and heat transport measurement\cite{R,Tan1} are interpreted in terms of nodes. Theoretically, an $s^{\pm}$-wave pairing symmetry (nodeless gap function with opposite signs of the order parameters for electron and hole pockets) has been suggested\cite{Lee}. Upon further doping, electron hole bands disappear, making the situation in the overdoped regime not so clear, though several forms of the order parameters have been proposed such as extended $s$, $d$, and $s + id$-wave pairing symmetries for heavily overdoped compounds\cite{Ch,Mai}. Therefore, the SC gap structure of the iron-based superconductors is not universal. It is quite different from that of cuprates in which almost all have a nodal pairing state\cite{Cup}. Such scattered pairing symmetries and various interpretations occur partly due to a sensitive dependence on measurement probes and material stoichiometry.

For clarifying this situation, we chose an electron-doped BaFe$_{2-x}$Ni$_{x}$As$_{2}$ system over the whole doping regime from its parent to an overdoped non-SC metal. The specific heat was systematically measured for 13 samples as marked by the arrows in Fig.\,1. The specific heat experiment has probably the best energy resolution of all experimental probes for distinguishing nodes in the gap. Near the absolute zero point, the bulk thermodynamic quantity is a sensitive probe of low-energy excitations of a complex quantum system. Low-energy excitations contain useful information about the nature of the ground state and can help to evaluate the presence or absence of nodes in the SC order parameter. For an $s$-wave superconductor, the electronic contribution to the specific heat is given by: C$_{el}$ $\propto \exp\frac{\Delta_{0}}{T}$, where $\Delta_{0}$ is the energy gap of the SC density of states (DOS). For a nodal gap symmetry, Sigrist and Ueda\cite{LN} showed that the line nodes in the gap are derived from the temperature dependence of the specific heat $\propto T^{2}$. Here, we present the experimental evidence for the crossover of the pairing symmetry from nodeless $s$-wave to node $d$-wave of BaFe$_{2-x}$Ni$_{x}$As$_{2}$ around $x = 0.15$. The results are in good agreement with the early experiments on the optimally doped compound\cite{W1,W2} and the penetration depth measurements on overdoped materials\cite{W4}.

To have full information on physical properties, we traced in Fig.\,1 the doping dependence of structural, magnetic, and SC properties of BaFe$_{2-x}$Ni$_{x}$As$_{2}$ single crystals as a function of Ni content. The undoped parent ($x$ = 0) undergoes both an antiferromagnetic (AFM) ordering and a structural transition from paramagnetic tetragonal to an AFM-ordered monoclinic phase at $T_{0}$ = 136(3)\,K with a first-order characteristic. Upon the Ni substitution, the sharp first-order structural/magnetic anomaly of the parent compound gradually broadens and shifts to lower temperatures\cite{PRB}. The commensurate static AFM order changes into transversely incommensurate short-range AFM order near optimal doping that coexists and competes with superconductivity. Upon further increasing the Ni concentration, the anomaly is no longer visible in the specific heat data, while superconductivity maintains at slightly lower temperatures, $i.e.,$ $T_{c}$ = 0.85\,K and 2.8\,K for $x$ = 0.22, and 0.20, respectively.

Figure\,2\textbf{a} illustrates the electronic specific-heat contribution of superconducting sample with $x$ = 0.096, obtained by subtracting the lattice specific heat from the raw data following the procedure given in ref.\,[24]. The entropy conservation required for a second-order phase transition is fulfilled. It is obvious that the SC transition at $T_\textup{c}$ is well pronounced showing a sharp jump in $C_\mathrm{el}$. We observed that superconductivity emerges in the magnetic phase for $x\leq$ 0.10 accompanied by increasing $T_{c}$. This increase is accompanied by a rise of the electronic coefficient $\gamma _{n}$ of the specific heat in the normal-state from 5.1\,mJ/mol K$^{2}$ at the undoped sample to about $24$\,mJ/mol K$^{2}$ to the optimal concentration, which we attribute to the closing of the SDW gap with Ni doping\cite{PRB}. Our data stresses that the decrease of $\gamma _{n}$ with increasing $x$ is linear for $x \geq 0.10$, while the residual electronic specific heat $\gamma _{r}$ linearly rises. This explains the increase of $T_{c}$ and $\Delta C_\textup{el}/ \gamma_\textup{n} T_\textup{c}$ with Ni doping in the underdoped region, implying that a larger fraction of the FS becomes available for superconductivity. However, the almost linear temperature dependence of $C_\mathrm{el}/T$ of the SC samples indicates that the specific heat data cannot be described by a single BCS gap. In order to illustrate this we show a theoretical BCS curve with $\Delta = 1.764\, k_{\mathrm{B}}T_\mathrm{c} = 2.23$~meV in Fig.\,2\textbf{a}. A precise description of the experimental data for the two-gaps $s$-wave model is obtained by using values of $\Delta _{1}(0)$ = 1.8\,k$_{B}T_{c}$ and $\Delta _{2}(0)$ = 0.74\,k$_{B}T_{c}$.

The magnetic field dependence of the specific heat through a vortex excitation in a mixed state is another excellent, sensitive independent testing method for identifying the gap structure of superconductors. It has been well demonstrated that $\gamma(H) \propto H$ for the isotropic $s$-wave because the specific heat in the vortex state is dominated by the contribution from the localized quasiparticle in the vortex core\cite{Vol}. It is clear from Fig.\,3 that the magnetic field enhances the low-temperature specific heat continuously, indicating the increase of the quasiparticle DOS at the FS induced by a magnetic field. The fact that the low-temperature specific heat data exhibits a linear behavior with magnetic field at low temperatures without any upturn indicates the absence of {Schottky-like} contributions in our samples. The data in Fig.\,3\textbf{a} roughly increases linearly with the applied magnetic field. Therefore, the observations of the temperature and field dependencies of the specific heat clearly show that the SC gap in the underdoped compounds is nodeless. The temperature dependence of the London penetration depth in the underdoped compounds provides strong evidence for a two-band $s$-wave order parameter\cite{PRB}. In addition, the lower critical field $H_{c1}$ values in BaFe$_{2-x}$Ni$_{x}$As$_{2}$ samples are measured by following the procedure described in ref.\,[24]. The corrected values of $H_{c1}$ are presented and associated with two-gap features. The $s^{++}$ wave is therefore identified to be the symmetry structure of the underdoped compounds. Additionally, the two-gap model also reproduces the electronic contribution to the specific heat of other doped samples well. The similar $s^{++}$ pairing symmetry has been observed in a phonon-mediated superconductor MgB$_{2}$ which has two $s$-wave energy gaps with the same phase in the $\sigma$ and $\pi$ bands\cite{Mg}.

However, the energy spectrum of quasiparticles $E_{k}$ in a singlet SC state can be given by $E_k=\sqrt{\xi^{2}_{k}+\Delta^{2}_{k}}$, where $\xi _{k}$ represents the energy of an electron relative to the chemical potential with momentum $k$. On the other hand, unconventional superconductivity is mostly characterized by the anisotropic SC gap function with nodes along certain directions in the momentum space. In the case of gap nodes, quasiparticles are generated to the largest extent in the vicinity of gap nodes. Moreover, the quasiparticle excitation spectrum takes the form $E_k=\hbar\sqrt{\upsilon_{\rm F}^{2}k_{\perp}^{2}+v_\Delta^{2}k_{\parallel}^2}$, where $k_{\perp}$ and $k_{\parallel}$ are wavevectors perpendicular and parallel to the FSs, respectively, $v_{\rm F}$ is the {\it renormalized} in-plane Fermi velocity at the position of the node, while $v_\Delta \approx \partial\Delta /\hbar \partial k $ is the slope of the gap at the node associated with the dispersion of the quasiparticles along the FS\cite{Vekhter2001}. This leads to a linear dependence of the DOS with energy. For the two-dimensional case it reads:
\begin{equation}\label{eqDOS}
N_{SC}(E) = \sum_{i}\frac{E}{\pi\hbar^2\upsilon_{\rm F}^i v_\Delta^i} ,
\end{equation}
where $i$ denotes the sum over all nodes. The observed $T^2$-behavior in Fig.\,2\textbf{c} suggests that this condition is fulfilled for the overdoped samples \emph{i.e}., $x$ = 0.18, 0.20, and 0.22. Using Eq.\ (\ref{eqDOS}) we also assumed that the energy gap at zero temperature, $\Delta_0$, is equal for all bands, while $\gamma_{el}$ is a {\it renormalized} normal-state Sommerfeld coefficient, being an average value over all FSs.

The SC order parameter $C/T \propto H^{0.5}$ holds, if the energy associated with the Doppler shift of $d$-wave quasiparticles in the vicinity of vortices $E_H= \hbar v_{\rm F}(\pi H/\Phi_0)^{0.5}$ is comparable to $E_T=k_bT$, where $\Phi_0$ is the flux quantum\cite{Vekhter2001}. According to our data shown in Fig.\,3\textbf{c} for $x$ = 0.18, the field dependence of $C/T$ follows a $H^{0.5}$-law which corresponds to the theoretical prediction\cite{Vekhter2001}. Therefore, the observed $T^{2}$ term together with the observed $H^{0.5}$ behavior of the specific heat in the SC state for the overdoped samples evidences $d$-wave superconductivity in almost all FS sheets. In that case the gap would necessarily have four line nodes that run vertically along the $c$ axis. In such a nodal structure, zero-energy nodal quasiparticles will conduct heat not only in the plane but also along the $c$ axis by an amount proportional to the $c$-axis dispersion of the FS\cite{J}. However, the nonlinear $\gamma(H)$ found in Fig.\,3\textbf{c} cannot be simply attributed to the multi-band effect as observed in Fig.\,3\textbf{a} in which $\gamma(H)$ is linear. Even if the multi-bands exist, the zero-field specific heat data shows a clear flattening in the low-temperature range corresponding to the weak excitation of quasiparticles for an $s$-wave superconductor. This is completely absent for the three compounds of $x$ = 0.18, 0.20, and 0.22 (Fig.\,2\textbf{c}).

In our investigated system, near the boundary between $s^{++}$ and $d$ states, we found an $s^{\pm}$ state. Theoretically\cite{Bang}, the $s^{\pm}$-pairing state with different gap sizes ($\Delta _{small} \neq  \Delta _{large}$) represents a strong field dependence $\gamma(H) \propto \sqrt{H} - H$. In addition, the most important virtual excitations that trigger the $s^{\pm}$-pairing are the AFM spin fluctuations. Although $s^{\pm}$-wave and $d$-wave are due to the Doppler shift effect by the supercurrent circulating around the vortices, the linear field dependence in the $s^{\pm}$-wave model is in contrast to the  $\sqrt{H}$ dependence\cite{Bang}. In Figs.\,2\textbf{b} and 3\textbf{b}, we overview theoretical calculations for both $s^{\pm}$ and $d$-wave approaches for $x$ = 0.15. It is obvious from the field dependence data that the $s^{\pm}$-wave model fits well with the field dependence which is taken as a signature for a nodal gap structure. In Fig.\,2\textbf{b}, we show the temperature dependence of the electronic contribution to the specific heat for $x$ = 0.15, obtained by subtracting the lattice specific heat as presented in ref.\,[24]. From an entropy-conserving construction a $T_\mathrm{c} = 13.12$~K was found. The specific heat jump at $T_{c}$ is 14.5 mJ/mol K$^{2}$ at zero field. The jump relative to the normal state specific heat is 0.83, which is much smaller than the BCS value 1.43. This indicates that the quasiparticles participating in the SC condensation experience strong elastic scattering because inelastic scattering usually enhances the jump ratio\cite{Bang}. However, the  value of the $\gamma _{r}/\gamma _{n}$ in the SC state for $x$ = 0.15 accounts for 26.6\%. This value also supports the presence of strong scatterers because weak scatterers are not capable of inducing states inside the SC gap of unconventional superconductors\cite{Bang}. However, in the case of $s^{\pm}$ superconductivity, $\gamma _{r}$ can be understood as arising from interband scattering. This could be induced by disorder due to Ni doping, which is pair-breaking for a sign-reversing order parameter\cite{Lee}.

A theoretical BCS curve with $\Delta = 1.764\, $k$_{\mathrm{B}}T_\mathrm{c} = 2.23$~meV is shown in Fig.\,2\textbf{b} with a small fraction $\gamma_\mathrm{res}/\gamma_n = 0.09$ of normal electrons. An $s^{\pm}$-wave pairing model was applied by combining the quasiclassical formalism with the first-principles calculation\cite{Nak}. Moreover, we used the $d$-wave approach to represent the generic behavior of a nodal SC gap state. More obviously, the temperature dependence presents a slightly concave-down behavior. This feature is characteristic of a typical multigap superconductor, in which a rather small SC gap coexists with large main gaps. This behavior suggests not only the existence of a small gap but also that of a medium-gap band\cite{Nak}. In Fig.\,2\textbf{b}, the $s^{\pm}$ approach is close to the weak-coupling single-band BCS result. One of the reasons is that the weighting of the small-gap band is small compared to the total one, \emph{i.e.}, the ratio $N_{3}$ /$N_{t}$ is 0.2, where $N_{t}$ is the total DOS\cite{Nak}. In addition, the gap-amplitude difference between the minimum $|\Delta_{3}|$ and the maximum $|\Delta_{2}|$ is not so large, where $|\Delta_{3}|$ /$|\Delta_{2}|$ $\approx$ 0.5. Moreover, $|\Delta_{2}|$/$T_{c}$ is found to be $\approx$ 2, which is bigger than the single-band value 1.76. Thus, the specific heat data from the temperature and magnetic field measurements is in favour of the sign changing multiband superconductivity of the $s^{\pm}$-wave state for $x$ = 0.15.

The reasoning for $s^{\pm}$ superconductivity may look quite straightforward and the $s^{\pm}$-pairing is favored when the antiferromagnetism is suppressed. To get $s^{\pm}$ superconductivity, one has to invoke some mechanism to enhance interpocket interactions\cite{s4}. However, the majority of researchers believe that $s^{\pm}$ is the right symmetry, even though the structure of $s^{\pm}$ SC order parameter has turned out to be more complex than originally thought. In strongly electron-doped Fe-based superconductors, $d$-wave superconductivity comes following $s^{\pm}$, but it emerges as the leading SC instability, for which the electron-hole interaction is reduced. The observation of a change of pairing symmetry in the same system upon doping would be unprecedented and is another reason why researchers are so excited about Fe-based superconductors\cite{s4}. Several groups have argued that a change from $s$-wave to $d$-wave would probably proceed through an intermediate state such as $s + is$ or $s + id$ symmetry, a complex order parameter that breaks time-reversal symmetry and $C_{4}$ lattice rotational symmetry\cite{s1,s2,s3,s4,s5}. In the current work the specific heat data was obtained on the samples with systematic doping variation by using a single bulk detection technique. Our investigated underdoped compounds are definitely consistent with nodeless multiband superconductivity. The large $T^{2}$ contribution and the $\sqrt{H}$ behavior at low temperatures for the overdoped samples provide evidence for the presence of line nodes in the SC gap. For the slightly overdoped level ($x$ = 0.15), we find that $s^{\pm}$ works well from both $T$ and $H$ dependence, indicating that its SC state is most compatible with the multiband $s^{\pm}$ wave.  However, at low temperatures the $s^{\pm}$ pairing symmetry could coexist with other SC phases, \emph{i.e.,} $t$ -type superconducting orders\cite{s1,Chub}, and the relative phase between the two is manifesting explicitly the breaking of the time-reversal symmetry. We thus obtained an interesting crossover of the pairing symmetry from $s^{++}$ to $d$-wave by passing an $s^{\pm}$ for a typical iron-pnictide family, which is believed to host unconventional pairing symmetry in these superconductors.

\begin{figure}
%\begin{center}
%\includegraphics[width=\columnwidth]{Fig1.pdf}\\
%\end{center}
\caption{\textbf{$|$ \,Gap symmetry crossover in electronic phase diagram of BaFe$_{2-x}$Ni$_{x}$As$_{2}$}. The phase diagram obtained from the magnetic and specific heat data, showing the suppression of the magnetic ($T_{N}$) and structural ($T_{S}$) phase transitions with an increasing Ni concentration and the appearance of the SC transitions. The arrows indicate the various doping levels studied in this work. Superconductivity exists in a dome-like region below the transition temperature $T_{c}$. With increasing $x$, $T_{c}$ rises to reach a maximal value of 20\,K at optimal doping. We show that the symmetry of the SC state changes from $s$-wave to $d$-wave (line nodes) and in between an $s^{\pm}$. The inset illustrates the sketch of the SC gap of $s$-wave vs. $d$-wave gap\cite{Cup}. The top row shows the momentum space representation in the regions of the Brillouin Zone having a (+) phase (pink) or (-) phase (blue). The bottom row illustrates the SC gap on a large Fermi surface. The $s$-wave has isotropic gap and it does not change sign. The $d$-wave gap changes sign four times as it goes around the Fermi surface and as a result the gap necessarily goes to zero at four nodes.}
\end{figure}

\begin{figure}
%\begin{center}
%\includegraphics[width=\columnwidth]{Fig2.pdf}\\
%\end{center}
\caption{\textbf{$|$ \,Temperature dependence of the specific heat of BaFe$_{2-x}$Ni$_{x}$As$_{2}$}. The measurements were performed in zero magnetic field and temperature down to $T$ = 70\,mK. The dashed lines in (\textbf{a}) for $x$=0.096 and (\textbf{b}) for $x$=0.15 represent the theoretical curves based on single-band weak-coupling BCS theory, while the dotted lines illustrate the $d$-wave approximation\cite{PRB}. The solid red line in (\textbf{a}) for $x$=0.096 indicates the two $s$-wave gap model. The  dot dashed line (\textbf{b}) for $x$ = 0.15 represents the $s^{\pm}$-wave model\cite{Nak}. The dotted lines in (\textbf{c}) for $x$ = 0.18, 0.20, and 0.22 (two samples) are the linear fits to the experimental data at low temperatures to extrapolate the residual linear terms at $T$ = 0. }
\end{figure}

\begin{figure}
%\begin{center}
%\includegraphics[width=\columnwidth]{Fig3.pdf}\\
%\end{center}
\caption{\textbf{$|$ \,Magnetic field dependence of the specific heat of BaFe$_{2-x}$Ni$_{x}$As$_{2}$}. The magnetic field was applied along the $c$ axis for the specific heat measurements, $i.e.$, $\emph{H} \parallel \emph{c}$. The solid line in (\textbf{a}) shows the linear fit to the experimental data for $x$=0.096, while the dotted line in (\textbf{c}) represents a fit to \emph{$\gamma$(H)} = \emph{A($\mu$$_{0}$H)$^{0.5}$} for $x=0.18$. For $x$=0.15 in (\textbf{b}), the dot dashed line is the theoretical fit with the $s^{\pm}$-wave model and the dotted line is the fit with the $d$-wave model, respectively. For an anisotropic gap, the specific heat is deviated from the field linear dependence.}
\end{figure}

\subsection{METHODS SUMMARY}

\subsection{}
\hspace{-3mm}
BaFe$_{2-x}$Ni$_{x}$As$_{2}$ ($x$ = 0, 0.03, 0.065, 0.085, 0.092, 0.096, 0.10, 0.12, 0.15, 0.18, 0.20, 0.22, and 0.25) single crystals were grown by the FeAs self-flux method, details for the  growth process and sample characterization were published elsewhere\cite{d,PRB}. The actual Ni level was determined to be 80\% of the nominal level $x$ through the inductively coupled plasma analysis of the as-grown single crystals. Magnetization measurements were performed by using a Quantum Design SC quantum interference magnetometer. The low-$T$ specific heat down to 0.4\,K was measured in its Physical Property Measurement System with the adiabatic thermal relaxation technique. Specific heat measurements were performed down to 70\,mK by using a heat-pulse technique within a dilution refrigerator along $H \parallel c$ up to $H$ = 9\,T.

\newpage

\vspace{16cm}
\begin{center}
\includegraphics[width=\columnwidth]{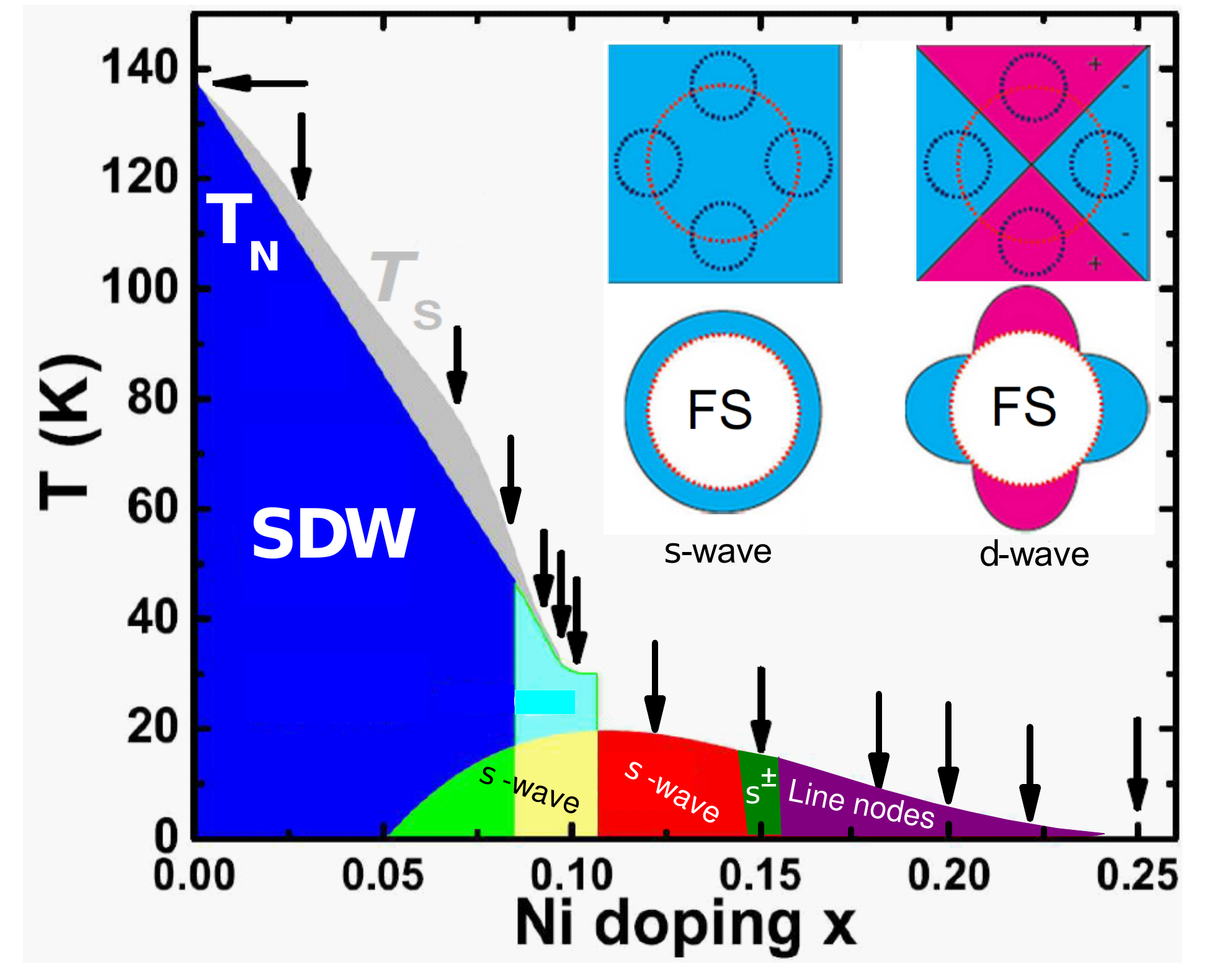}
{\item Fig. 1.}
\end{center}

\newpage

\begin{center}
\includegraphics[width=\columnwidth]{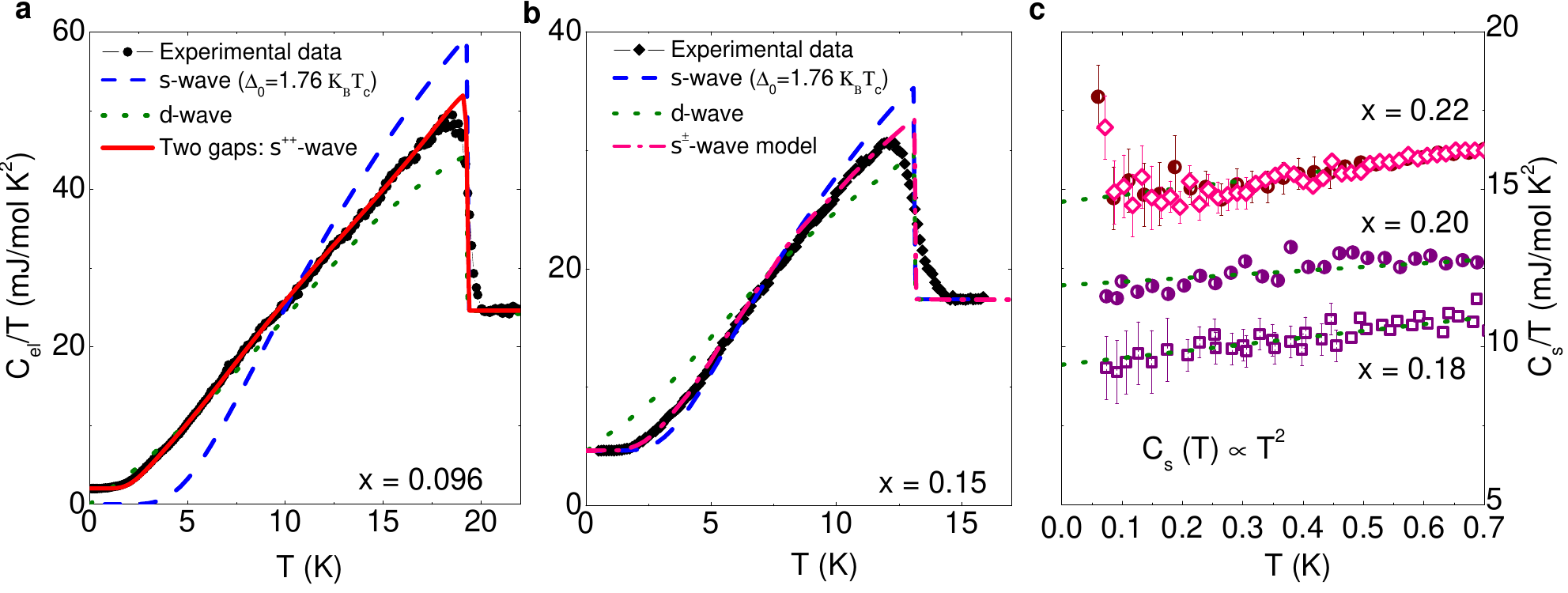}
{\item Fig. 2.}
\end{center}

\vspace{-15cm}
\begin{center}
\includegraphics[width=\columnwidth]{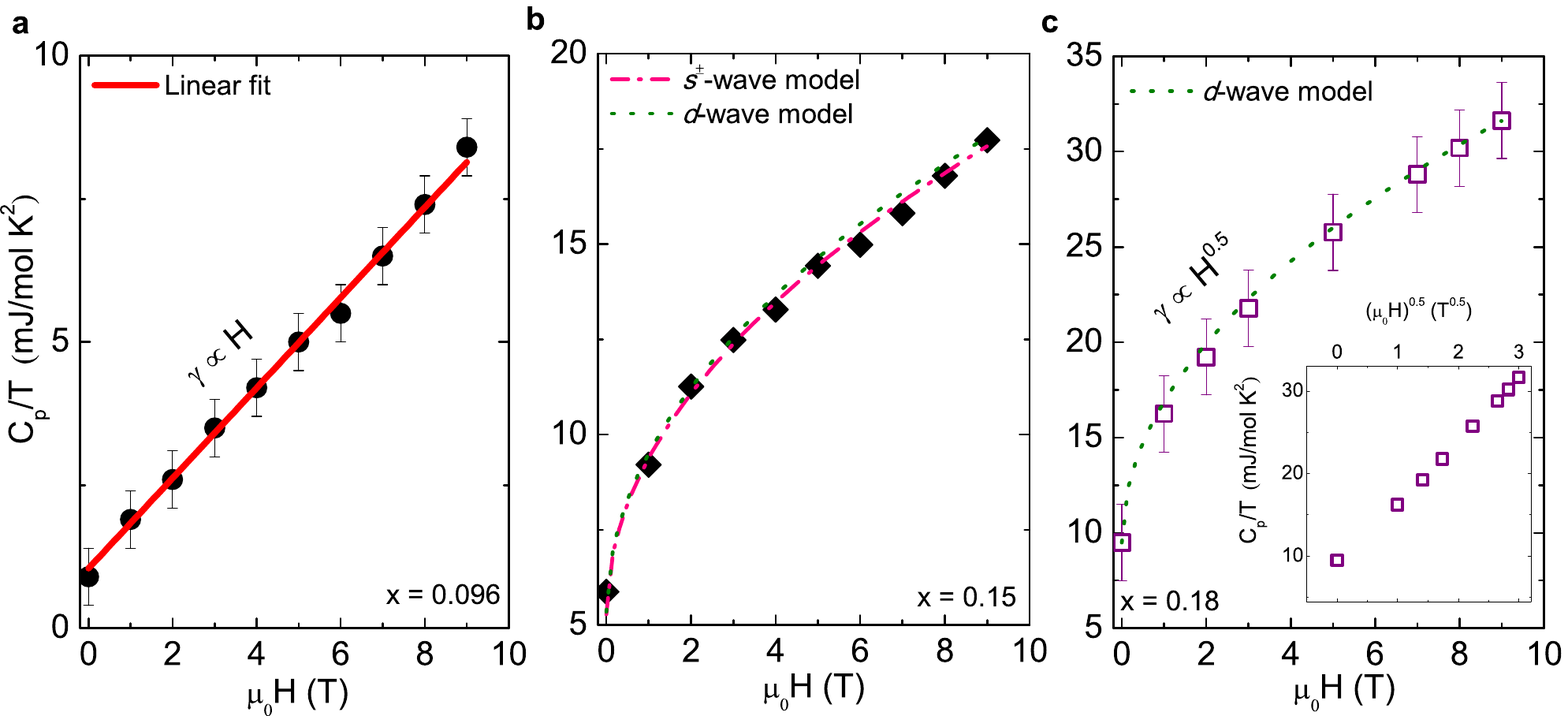}
{\item Fig. 3.}
\end{center}

\begin{addendum}
 \item We appreciate the useful discussions with R. M. Fernandes, Donglai Feng, Igor Mazin, Juan Jiang, Goran Karapetrov, Helge Rosner, Christoph Geibel, Alexander Vasiliev, and Ruediger Klingeler. Z.H. and J.Z. acknowledge the support from the Shanghai Pujiang Scholar program (No.13PJ1401100). The works in IOP of CAS are supported by the Natural Science Foundation of China (Nos. 11374011 and 91221303), the Ministry of Science and Technology of China (973 projects: Grants Nos. 2011CBA00110 and 2012CB821400), and The Strategic Priority Research Program (B) of CAS (Grant No. XDB07020300). The work at Rice is supported by National Science Foundation Grant No. DMR-1362219 and the Robert A. Welch Foundation Grant No. C-1839.

\item[Author Information] Correspondence and requests for materials should be addressed to X.J.C. (xjchen@hpstar.ac.cn).

\end{addendum}

\end{document}